\documentclass[twocolumn,english,aps, prd,nofootinbib,  superscriptaddress, preprintnumbers,floatfix]{revtex4-2}
\usepackage[T1]{fontenc}
\usepackage{babel}
\usepackage{mathrsfs}
\usepackage{bm,amsmath}
\usepackage{amssymb}

\usepackage{graphicx}
\usepackage{hyperref}
\hypersetup{pdftex,colorlinks=true,linkcolor=blue,citecolor=blue,menucolor=black,urlcolor=blue,filecolor=blue}

\usepackage{slashed}
\usepackage{gensymb}
\usepackage{bbold}
\usepackage{multirow}

\usepackage{amsfonts}
\usepackage{makecell}
\usepackage[dvipsnames]{xcolor}
\usepackage[normalem]{ulem}
\usepackage{longtable}
\usepackage{array}
\usepackage{float}
\usepackage{subfigure}
\usepackage{orcidlink}

\usepackage{kpfonts}

\newcommand{\mllb}{m_{l\bar l}}

\newcommand*{\Rom}[1]{\uppercase\expandafter{\romannumeral #1\relax}}
\newcommand{\pvec}[1]{\vec{#1}\mkern2mu\vphantom{#1}}

\newcommand{\itp}{\affiliation{CAS Key Laboratory of Theoretical Physics, Institute of Theoretical Physics,\\
Chinese Academy of Sciences, Beijing 100190, China}}

\newcommand{\ucas}{\affiliation{School of Physical Sciences, University of Chinese Academy of Sciences, Beijing 100049, China}}

\newcommand{\peng}{\affiliation{Peng Huanwu Collaborative Center for Research and Education, Beihang University, Beijing 100191, China}}

\newcommand{\hiskp}{\affiliation{Helmholtz-Institut f\"ur Strahlen- und Kernphysik and Bethe Center for Theoretical Physics,\\ Universit\"at Bonn, D-53115 Bonn, Germany}}

\newcommand{\fzj}{\affiliation{Institute for Advanced Simulation,  Forschungszentrum J\"ulich, D-52425 J\"ulich, Germany}}

\newcommand{\tbilisi}{\affiliation{Tbilisi State University, 0186 Tbilisi, Georgia}}

\begin{document}
\title{The Proton Charge Radius from Dimuon Photoproduction off the Proton}

\author{Yong-Hui Lin\orcidlink{0000-0001-8800-9437}}\email{yonghui@hiskp.uni-bonn.de}\hiskp

\author{Feng-Kun Guo\orcidlink{0000-0002-2919-2064}}\email{fkguo@itp.ac.cn}\itp\ucas\peng

\author{Ulf-G.~Mei{\ss}ner\orcidlink{0000-0003-1254-442X}}\email{meissner@hiskp.uni-bonn.de}\hiskp\fzj\tbilisi

\begin{abstract}
	We investigate the feasibility of measuring the proton charge radius through dimuon photoproduction off a proton target.
 Our findings indicate that the Bethe-Heitler mechanism, which dominates at small momentum transfers, allows for an extraction of the proton electromagnetic form factors in the extremely low $Q^2$ region below $10^{-3}$~GeV$^2$ in the spacelike region, when the incident photon beam energy exceeds several hundred MeV. The optimal kinematical region and a sensitivity study of the proton charge radius from dimuon photoproduction are presented. Such a measurement is expected to provide an alternative to the elastic muon-proton scattering measurements such as MUSE at PSI and AMBER at CERN.
\end{abstract}
\maketitle

\section{Introduction}
The determination of the proton charge radius with high accuracy has promoted significant efforts from both the theoretical and experimental communities since 2010, when the first muonic hydrogen spectroscopic measurement of the proton charge radius was reported~\cite{Pohl:2010zza}. 
Incorporating the updated reevaluation~\cite{Antognini:2013txn} revealed an unexpected 5.6$\sigma$ discrepancy, in that
the radius from muonic hydrogen turned out to be significantly
smaller than the radius given by the CODATA group averaging
over elastic electron-proton scattering and electronic hydrogen spectroscopic measurements~\cite{A1:2010nsl,CODATA:2014}. 
To date, the smaller radius about 0.84~fm from the muonic hydrogen spectroscopic measurement has been confirmed by several independent determinations: two electronic hydrogen spectroscopic results, one in 2017~\cite{Beyer:2017gug} from a measurement of the 2S$\to$4P transition combined with measurements of the 1S$\to$2S transition~\cite{Parthey:2011lfa,Matveev:2013orb} and the other in 2019~\cite{Bezginov:2019mdi} from a direct measurement of the 2S$\to$2P Lamb shift,
an electron-proton scattering measurement in 2019~\cite{Xiong:2019umf}, and from the theoretical side, a dispersion theoretical analysis of the full-range form factor data in 2022~\cite{Lin:2021xrc}, as well as precise lattice QCD calculations recently~\cite{Djukanovic:2023jag,Djukanovic:2023beb,Tsuji:2023llh}. 
Given this progress, the latest recommended value of the proton charge radius by the CODATA group has been updated to 0.84075(64)~fm~\cite{codata2024}, consistent with the muonic hydrogen spectroscopic value. 
For reviews on the progress in the determination of the proton charge radius, we refer to~\cite{Karr:2020wgh,Gao:2021sml,Peset:2021iul,Antognini:2022xoo}. 

From this, one might conclude that the proton radius ``puzzle'' has been solved, shifting the focus from a puzzle to precision~\cite{Hammer:2019uab}. However, some tension still exists, as one recent measurement in 2022 of the hydrogen $2 \mathrm{S}_{1 / 2}$--$8 \mathrm{D}_{5 / 2}$ transition~\cite{Brandt:2021yor} combined with the $1 \mathrm{S}_{1 / 2}$--$2 \mathrm{S}_{1 / 2}$ value measured in Ref.~\cite{Parthey:2011lfa} led to a value of 0.8584(51)~fm, showing a 3.1$\sigma$ deviation from the above quoted CODATA value.
Furthermore, it is crucial to address the missing piece in the determination of the proton charge radius through muon-proton scattering (such a determination will be called the muon-proton scattering value of the proton charge radius). This is also essential for testing the lepton flavor universality, a cornerstone of the Standard Model.

There are currently two projects  proposed to implement elastic muon-proton scattering measurements of the proton charge radius, that is, MUSE at PSI aiming at accessing the $Q^2$ range from 0.002 to 0.07~GeV$^2$~\cite{Downie:2014qna} and AMBER at CERN~\cite{Adams:2018pwt} with $0.001\ {\rm GeV^2}< Q^2<0.02\ {\rm GeV^2}$. One of the challenges in the muon-proton scattering experiment is to construct a clean muon beam, as the commonly used secondary muon beam is always contaminated by electrons and pions~\cite{MUSE:2013uhu}. 

\begin{figure}[t]
	\centering
	\includegraphics*[width=0.45\textwidth,angle=0]{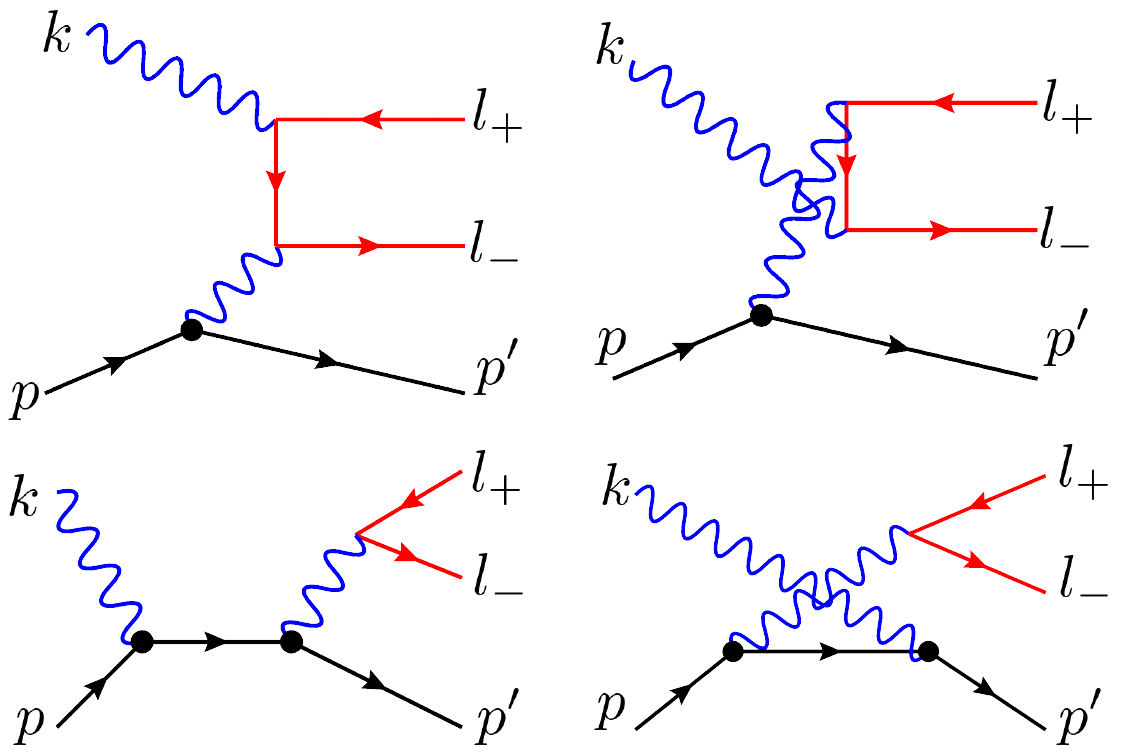}
	\caption{
		Diagrams for the $\gamma(k)p(p)\to p(p^\prime) \mu^+(l_+)\mu^-(l_-)$ process. The upper panel shows the Bethe-Heitler mechanism and the lower panel shows the Compton scattering mechanism.
	}
	\label{fig:diagram}
\end{figure}

In this work, we perform a systematic feasibility study on extracting the muon-proton scattering value of 
the proton charge radius in the dimuon photoproduction on a proton target. Both the Bethe-Heitler (BH) process (upper panel in Fig.~\ref{fig:diagram}) and the timelike Compton scattering (TCS) process (lower panel in Fig.~\ref{fig:diagram}) contribute to  dimuon photoproduction\footnote{All Feynman diagrams in this work were drawn using FeynGame-2.1~\cite{Harlander:2024qbn}.}. Notice that the BH process has the same hadronic operator in its amplitude as elastic muon-proton scattering, making it a viable alternative to access the muon-proton scattering value of the proton charge radius without the need for constructing a muon beam. Similar reactions were proposed to test the lepton universality when extracting the proton charge form factor~\cite{Pauk:2015oaa}, to measure the deuteron charge radius~\cite{Carlson:2018ksu}, and to study the parton structure of the proton~\cite{Berger:2001xd,Boer:2015fwa}.

\section{Formalism for dimuon photoproduction on a proton}
To access the proton charge radius, i.e., to extract the low-$Q^2$ proton electromagnetic form factors (EMFFs) through the BH process, it is essential to identify the optimal kinematic region where the background TCS process is significantly suppressed. To this end, we present some necessary formulas for calculating the contributions of the BH and TCS processes to the dimuon photoproduction off a proton $\gamma(k)+ p(p) \to p(p^\prime)+ \mu^+(l_+)+\mu^-(l_-)$, see also Refs.~\cite{Berger:2001xd,Boer:2015fwa,Pauk:2015oaa,Heller:2018ypa,Heller:2019dyv,Heller:2020lnm}.
The general expression for the differential cross section of a $2\to3$ reaction with one massless particle in the initial state is given by 
\begin{align}\label{eq:dcs0}
	&\frac{d\sigma}{d m_{l\bar l}d\cos\theta^* d\phi^* d\cos\theta^\prime}=\frac{1}{16(2\pi)^4}\notag\\
	&\phantom{xx}\times\frac{\sqrt{\lambda(m_{l\bar l}^2,m_\mu^2,m_\mu^2)}}{m_{l \bar l}}\frac{\sqrt{\lambda(s,m_p^2,m_{l\bar l}^2)}}{4s (s-m_p^2)}\frac14\sum_{\rm spins} |{\cal M}|^2,
\end{align}
where $\lambda(x,y,z)=x^2+y^2+z^2-2(xy+yz+zx)$ is K\"{a}ll\'{e}n's triangle function, $s=(p+k)^2=m_p^2+2m_p E_\gamma$, and $m_{l \bar l}^2=(l_++l_-)^2$, with $m_p$ the proton mass, $m_\mu$ the lepton mass and $E_\gamma$ the photon energy in the \textit{lab} frame where the target proton is at rest. The polar and azimuthal angles ($\theta^*$, $\phi^*$) describe the direction of the muon momentum $l_-$ in the $l\bar l$ \textit{c.m.} frame, while the angles ($\theta^\prime$, $\phi^\prime$) represent the direction of the scattered proton momentum $p^\prime$ in the $\gamma p$ \textit{c.m.} frame. 

The amplitude for the diagrams in Fig.~\ref{fig:diagram} can be written as 
\begin{equation}
	i{\cal M}=\frac{-ig_{\mu\nu}}{t}L_{\rm BH}^\mu H_{\rm BH}^\nu +\frac{-ig_{\mu\nu}}{m_{l\bar l}^2}L_{\rm TCS}^\mu H_{\rm TCS}^\nu,
\end{equation}
where $t=(p^\prime-p)^2$ is the transfer momentum squared of the exchanged photon for the BH mechanism, $L_{\rm TCS/BH}$ denotes the leptonic operator consisting of the fundamental quantum electrodynamics (QED) vertices of the lepton, while $H_{\rm TCS/BH}$ represents the hadronic operator, incorporating the non-pointlike structure of the proton as seen by the photon probe that is described conventionally by two independent form factors, $F_1$  and $F_2$, known as the Dirac and Pauli form factors, respectively. The leptonic and hadronic tensors are:
\begin{align}
\begin{split}
    L_{\rm BH}^\mu=&\,\frac{i e^2}{(l_+-k)^2-m_\mu^2} \gamma^\mu(\slashed{l}_+-\slashed{k}-m_\mu)\slashed{\epsilon}(k) \\
	&+\frac{i e^2}{(-l_-+k)^2-m_\mu^2} \slashed{\epsilon}(k)(-\slashed{l}_-+\slashed{k}-m_\mu)\gamma^\mu,\\
	H_{\rm BH}^\mu= &\, i e \bar{u}(p^\prime)\Gamma_{\rm BH}^\mu u(p)
\end{split}
\end{align}
for the BH process, and
\begin{align}
\begin{split}
    L_{\rm TCS}^\mu=&\, -i e \bar{u}(l_-)\gamma^\mu v(l_+),\\
	H_{\rm TCS}^\mu=&\,\frac{-i e^2\epsilon_\nu(k)}{(p+k)^2-m_p^2}\bar{u}(p^\prime)\Gamma_{{\rm TCS},f}^\mu(\slashed{p}+\slashed{k}+m_p)\Gamma_{{\rm TCS},i}^\nu u(p)\\
	&+\frac{-i e^2\epsilon_\nu(k)}{(p^\prime-k)^2-m_p^2}\bar{u}(p^\prime)\Gamma_{{\rm TCS},i}^\nu(\slashed{p}^\prime-\slashed{k}+m_p)\Gamma_{{\rm TCS},f}^\mu u(p)
\end{split}
\end{align}
for the TCS process, with $\Gamma_{\rm BH}$, $\Gamma_{{\rm TCS},i/f}$ the $\gamma p p$ vertices. 
Employing the on-shell assumption for $\Gamma_{{\rm TCS},i/f}$, we have
\begin{align}
	\Gamma_{\rm BH}^\mu&=\gamma^\mu F_1(t)+\frac{i\sigma^{\mu\nu}(p^\prime-p)_\nu}{2 m_p}F_2(t),\\
	\Gamma_{{\rm TCS},i}^\mu&=\gamma^\mu+\frac{i\sigma^{\mu\nu}k_\nu}{2 m_p}\kappa_p,\\
	\Gamma_{{\rm TCS},f}^\mu&=\gamma^\mu F_1(m_{l\bar l}^2)-\frac{i\sigma^{\mu\nu}(l_++l_-)_\nu}{2 m_p}F_2(m_{l\bar l}^2)\label{eq:onshell}.
\end{align}
Here, $\kappa_p=1.793$ is the anomalous magnetic moment of the proton~\cite{ParticleDataGroup:2024}, which provides the normalization for the Pauli form factor $F_2$. {The uncertainty associated with the on-shell approximation for the TCS process is completely negligible for our purpose as discussed in detail in  Appendix~\ref{app:offshell}.} Further, the so-called Sachs form factors are
\begin{equation}
    G_E=F_1-\tau F_2, \quad G_M=F_1+F_2,
\end{equation}
where $\tau=-t/(4m_p^2)$ for the BH process and $\tau=-\mllb^2/(4m_p^2)$ for the TCS process.

Note that the EMFFs of the proton in the spacelike region, at the photon point, and in the timelike region are accessed by the $\Gamma_{\rm BH}$, $\Gamma_{{\rm TCS},i}$ and $\Gamma_{{\rm TCS},f}$, respectively, corresponding to the momentum transfer squared $t<0$, $k^2=0$, and $m_{l\bar l}^2>0$. Our focus is on the experimental strategy for extracting $\Gamma_{\rm BH}$. The same conventions as in Refs.~\cite{Pauk:2015oaa,Carlson:2018ksu} are used for choosing the two kinematic variables ($t$ and $m_{l\bar l}^2$) to investigate the differential cross section.  
To be concrete, we investigate the differential cross section of $\gamma p \to p\mu^+\mu^-$ as a function of the momentum transfer squared $t$ and the invariant mass squared of the lepton pair $m_{l\bar l}^2$, with the lepton angles $\theta^*$ and $\phi^*$ integrated out. This implies that we only need to detect the momentum and angle of the recoiling proton (c.f. Eq.~\eqref{eq:dcs0}), and furthermore, its scattering angle can also be fixed, as we shall show. 
This impressive feature of fixing the scattering angle in detection makes the dimuon photoproduction on a proton significantly more advantageous than the elastic muon-proton scattering for experimental design and implementation, despite the
suppression with $\alpha$ in the cross section. The following kinematic relations are useful: 
\begin{align}
	|\pvec{p}^\prime|^{\textit{lab}}&=2 m_p\sqrt{\tau(1+\tau)},\label{eq:plab}\\
	\cos\theta_{p^\prime}^{\textit{lab}}&=\frac{m_{l\bar l}^2+2(s+m_p^2)\tau}{2(s-m_p^2)\sqrt{\tau(1+\tau)}},\label{eq:anglelab}\\
    \cos\theta^\prime&=\frac{m_{l\bar l}^2(s+m_p^2)-2s t-(s-m_p^2)^2}{(s-m_p^2)\sqrt{\lambda(s,m_p^2,m_{l\bar l}^2)}},
\end{align}
with $\tau=-t/(4m_p^2)$, where $|\pvec{p}^\prime|^{\textit{lab}}$ and $\theta_{p^\prime}^{\textit{lab}}$ are the magnitude and angle of the momentum of the recoiling proton in the {\em lab} frame, respectively. 

\begin{figure}[tb]
	\centering
	\includegraphics*[width=0.45\textwidth,angle=0]{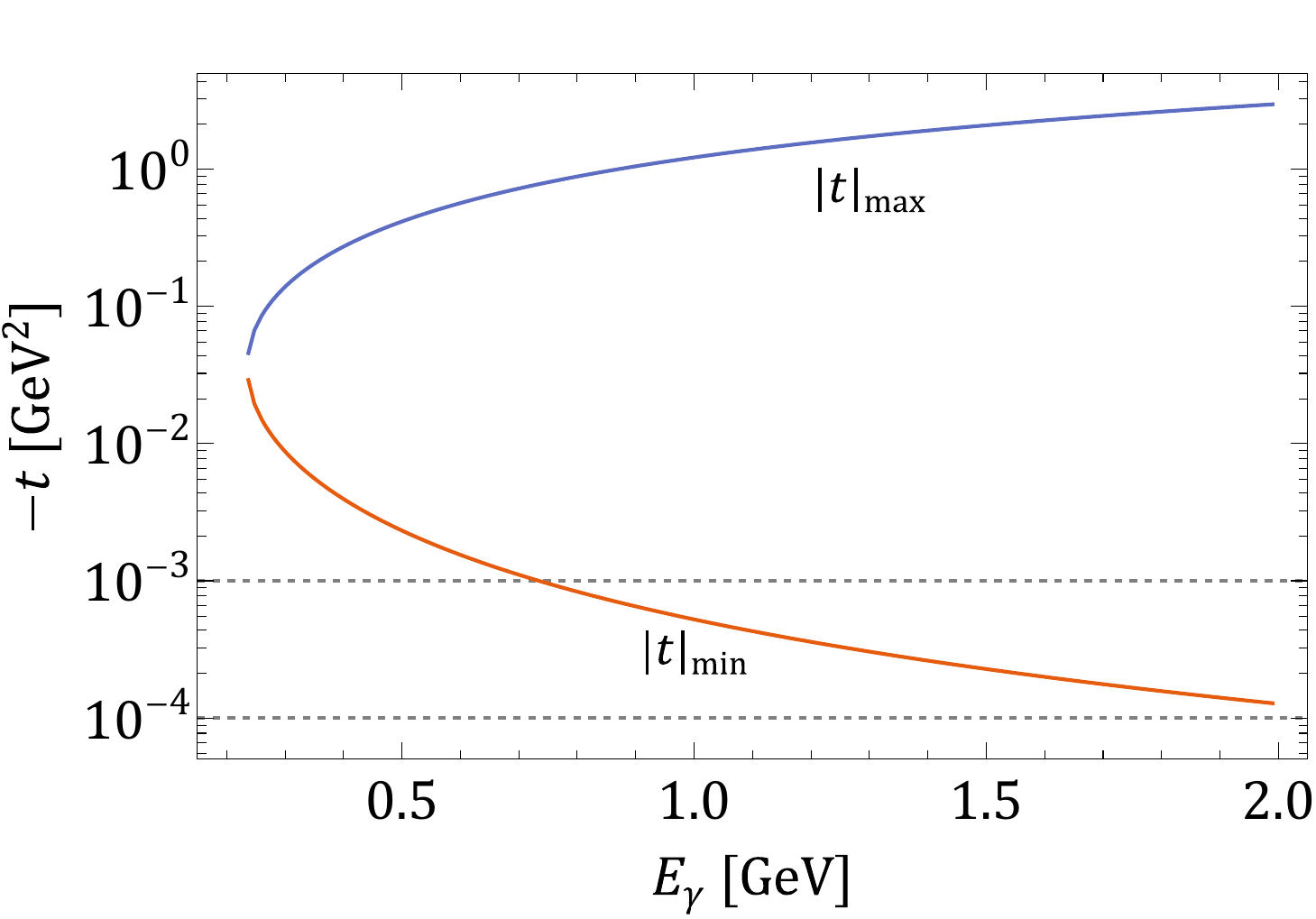}
	\caption{
		The kinematically allowed $t$ range for various incident photon \textit{lab} energies.
	}
	\label{fig:trange}
\end{figure}

In Fig.~\ref{fig:trange}, we plot the allowed range of the momentum transfer squared $t$ depending on the incoming photon \textit{lab} energy $E_\gamma$. We find that when the photon \textit{lab} energy exceeds 0.8~GeV, the minimal value of the momentum transfer squared $Q^2=-t$ can reach $10^{-3}\ {\rm GeV^2}$, the lowest value accessible by AMBER. Then we focus on exploring the competition between the TCS and BH processes for photon \textit{lab} energy above 0.8~GeV. 

\begin{figure}[tb]
	\centering
	\includegraphics*[width=0.45\textwidth,angle=0]{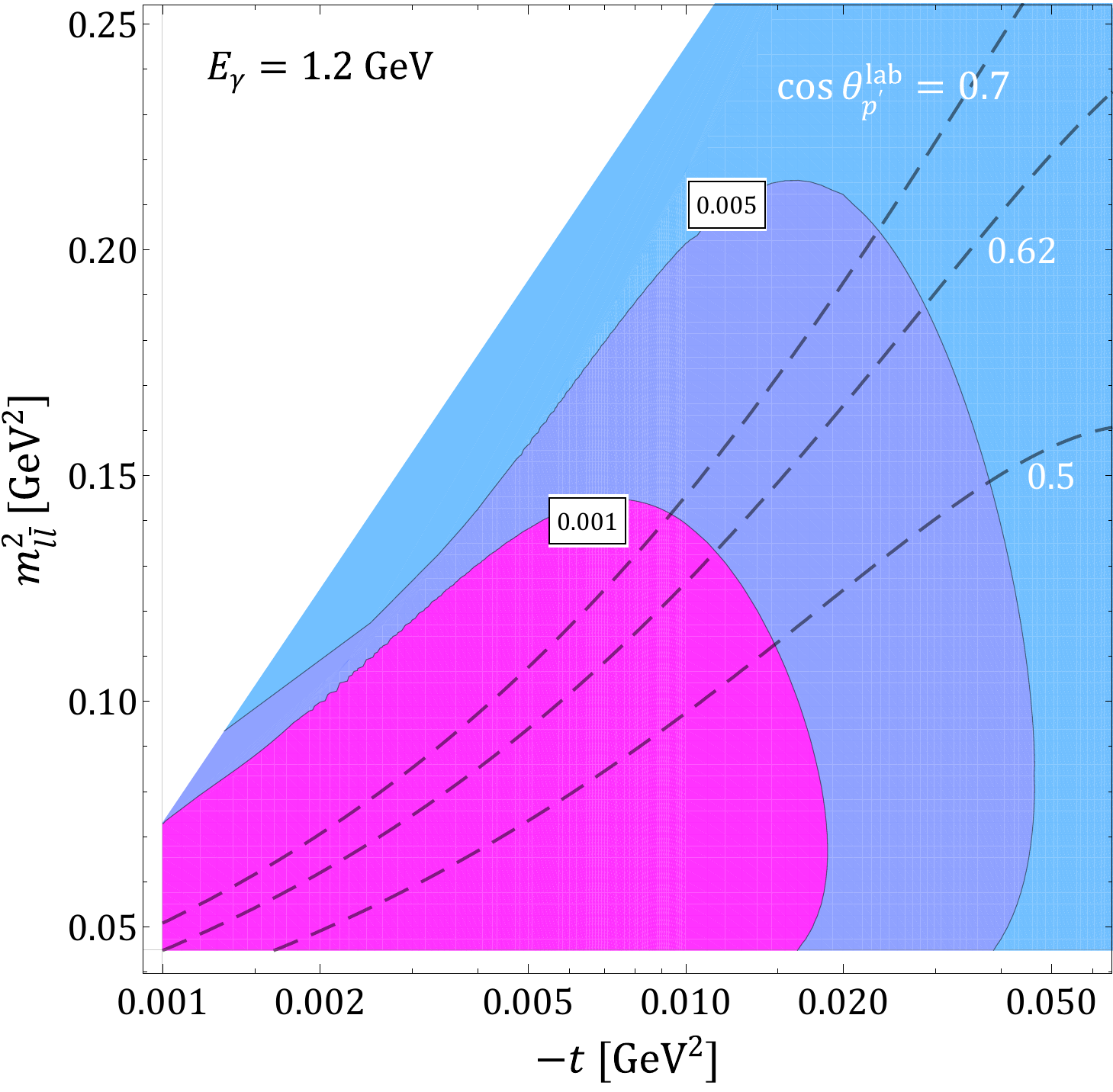}
	\caption{Counter plot of the ratio of differential cross sections from the TCS and BH processes,
		$d\sigma_{\rm TCS}/d\sigma_{\rm BH}$, at the photon \textit{lab} energy $E_\gamma=1.2\ {\rm GeV}$. The dashed lines show the value of $\cos \theta_{p^{\prime}}^{\textit{lab}}$.
	}
	\label{fig:kinematic}
\end{figure}

Figure~\ref{fig:kinematic} shows the contour plot of the ratio of differential cross sections from the TCS and BH processes, $d\sigma_{\rm TCS}/d\sigma_{\rm BH}$, in the $(-t,m_{l\bar l}^2)$ plane for an incident photon \textit{lab} energy of 1.2~GeV. The magenta area, which occupies a large fraction of the small-$Q^2$ and low-$m_{l\bar l}^2$ region, indicates where $d\sigma_{\rm TCS}/d\sigma_{\rm BH}<0.001$. It makes measuring the proton charge radius from the reaction $\gamma p \to p\mu^+\mu^-$ through the BH process in this kinematic region feasible. 

Notice that Eq.~\eqref{eq:plab} establishes a one-to-one correspondence between the \textit{lab} momentum of the proton and the spacelike momentum transfer squared $t$. Furthermore, for a given value of $t$, the invariant mass squared of the lepton pair $m_{l\bar l}^2$ can be obtained from the angle of the recoiling proton in the \textit{lab} frame, $\theta_{p^\prime}^{\textit{lab}}$, via the kinematic relation of Eq.~\eqref{eq:anglelab}. This allows for the extraction of 
the proton electromagnetic form factors from experimental cross sections at various $|\pvec{p}^\prime|^{\textit{lab}}$ points with fixed \textit{lab} angle $\theta_{p^\prime}^{\textit{lab}}$. 

The dashed curves shown in Fig.~\ref{fig:kinematic} represent various contours for different fixed values of the \textit{lab} angle $\theta_{p^\prime}^{\textit{lab}}$ of the recoiling proton within the specified kinematic region. For $\cos\theta_{p^\prime}^{\textit{lab}}>0.62$, the lowest value of $10^{-3}\ {\rm GeV^2}$ for $-t$ accessible by AMBER can be reached. In the remaining analysis, we restrict ourselves to the proposed optimal kinematic setup, i.e., $0.001\ {\rm GeV^2}<-t<0.02\ {\rm GeV^2}$ with $E_\gamma=1.2\ {\rm GeV}$ and $\cos\theta_{p^\prime}^{\textit{lab}}=0.7$.

\section{Sensitivity to the proton charge radius}

It is instructive to study the sensitivity of the cross section of the proposed reaction to the proton charge radius, which is necessary for future experimental investigations. We employ the same strategy as in our previous work~\cite{Lin:2023qnv}, that is, fitting the proton EMFFs to the Monte Carlo pseudodata of the $\gamma p \to p\mu^+\mu^-$ cross sections generated using the von Neumann rejection method, adhering to a specified distribution. As depicted in Fig.~\ref{fig:kinematic}, within the proposed optimal kinematic region, the differential cross section of dimuon photoproduction on the proton can be accurately described by the BH process, with an uncertainty smaller than 0.1\%. A very compact expression for the BH differential cross section can be found in Refs.~\cite{Pauk:2015oaa,Heller:2019dyv}. We present the expression in our notation in Appendix~\ref{app:BHxs}. 

To proceed, we use as input the dipole electric form factor with $r_E^p=0.840\ {\rm fm}$ and Kelly's magnetic form factor~\cite{Kelly:2004hm} to produce a sample distribution of the BH differential cross section. 
The cross section, after integrating $-t$ over the range from 0.001~GeV$^2$ to 0.02~GeV$^2$, is estimated to be 124~nb ($\sim {\cal O}(100)$~nb) using this prescription for the proton EMFFs. 
Considering the experimental setup where a photon beam from gamma-ray sources with a flux of $10^7$~photons/s (e.g., ELSA
at Bonn~\cite{Hillert:2006yb}, MAMI at Mainz~\cite{Ostrick:2011zza},
GRAAL at the European Synchrotron Radiation Facility~\cite{Nedorezov:2012zz} and LEPS at SPring-8~\cite{Muramatsu:2020odu} have gamma beams above 1~GeV with $10^6$-$10^7$ photon flux available; {see also Ref.~\cite{Yu:2018cfr} for a proposal of generating high brilliance $\gamma$ rays up to 1~GeV using a 10-PW laser}) 
impinges on a 1 m long Time-Projection Chamber (TPC) target\footnote{For the measurement of low-energic recoil protons, a TPC-type active target is necessary~\cite{Adams:2018pwt,COMPASSAMBERworkinggroup:2019amp}.} filled with pressurized hydrogen gas up to 20 bar, approximately $5\times 10^6$ events of the desired BH signal would be available after several months of data collection.

\begin{figure}[t]
	\centering
	\includegraphics*[width=0.48\textwidth,angle=0]{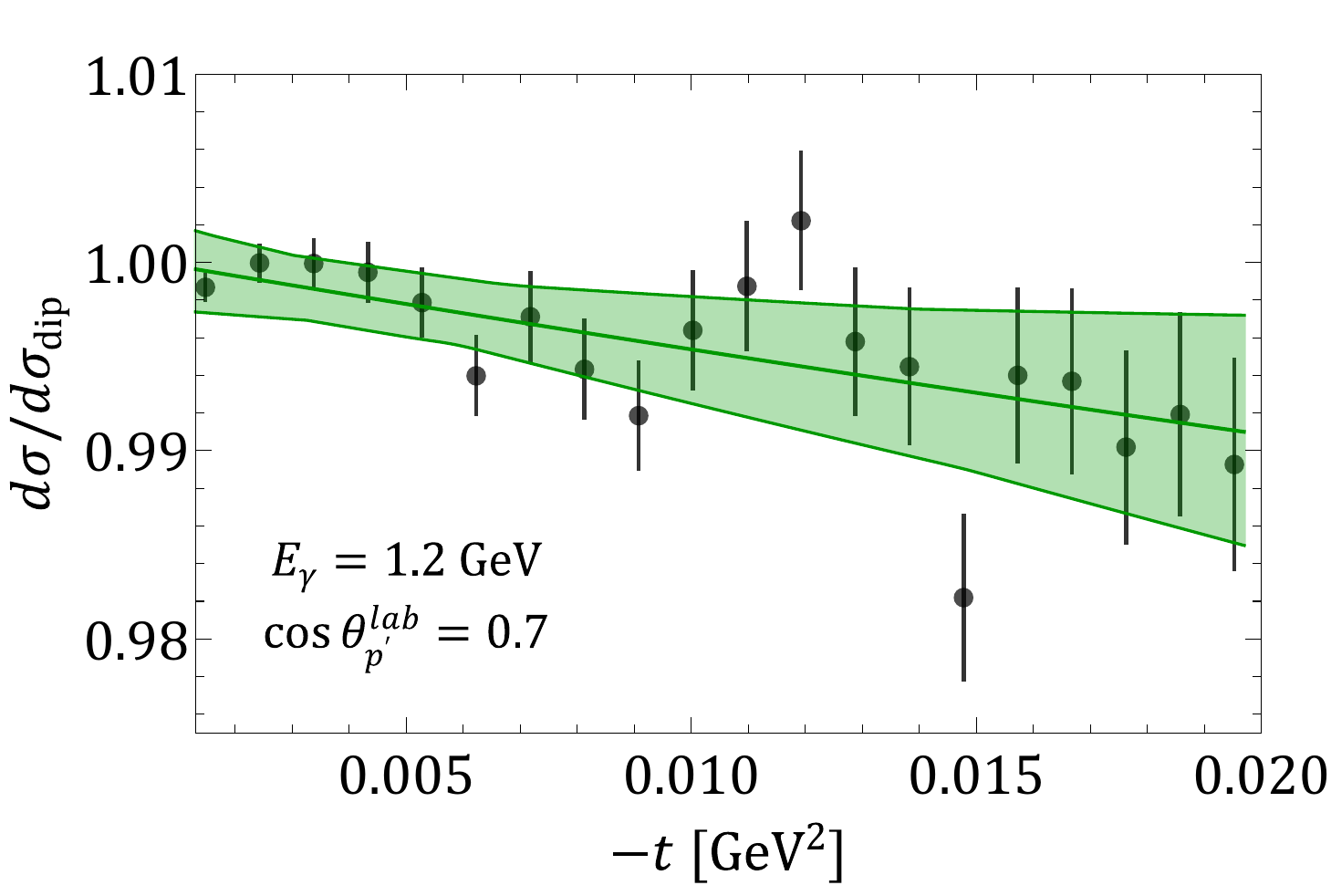}
	\caption{
		Fits to $5\times 10^6$ synthetic Monte Carlo events that adhere to the sample distribution of the BH differential cross sections, see the descriptions in the text. The green line and band show the best fit with the dispersion-theoretical parameterization for the proton EMFFs and the corresponding uncertainty, estimated using a Bayesian technique.
	}
	\label{fig:sensitivity}
\end{figure}

In Fig.~\ref{fig:sensitivity} we display the fit results using the dispersion relation parameterization of the proton EMFFs (for details, we refer to the recent review~\cite{Lin:2021umz}) to $5\times 10^6$ Monte Carlo events divided into 20 bins. 
The plot shows the cross section normalized to results from the standard dipole form factors for the proton, $\sigma_{\rm dip}$, that is, $G_E=G_M/(1+\kappa_p)=G_D=(1-t/(0.71\ {\rm GeV^2}))^{-2}$. 
The extracted proton charge radius is $0.848(8)$~fm with the uncertainty, which is propagated from the pseudodata, estimated using a Bayesian technique as detailed in Refs.~\cite{Lin:2021umz,Schindler:2008fh,Wesolowski:2015fqa}.
In this study, we implement the delayed rejection adaptive
metropolis algorithm (DRAM)~\cite{2020arXiv200914229S} to do the Bayesian simulation.
We find that a cross section measurement with a 0.5\% uncertainty will allow for an extraction of the muon-proton scattering value of the proton charge radius at the 1\% level.

Additionally, the full leading order QED radiative corrections to both the TCS and BH processes in the $\gamma p \to p\mu^+\mu^-$ reaction have been investigated recently by Refs.~\cite{Heller:2018ypa,Heller:2019dyv,Heller:2020lnm}.
When real data become available, such radiative corrections can also be included in the analysis to extract the proton charge radius. The radius obtained in this way can be directly compared with the upcoming measurements from the elastic muon-proton scattering experiments. 

\section{Summary and prospect}
In this work, we present a systematic study of the dimuon photoproduction off a proton with the aim of extracting the muon-proton scattering value of the proton charge radius, which has not been measured yet. 
We have shown that the Bethe-Heitler process dominates in the small momentum transfer region. 
The optimal kinematical setup to extract the proton charge radius is proposed to be $0.001\ {\rm GeV^2}<-t<0.02\ {\rm GeV^2}$ with $E_\gamma=1.2\ {\rm GeV}$ and $\cos\theta_{p^\prime}^{\textit{lab}}=0.7$, where the background contribution from the timelike Compton scattering process is smaller than 0.1\%. 
Moreover, with a Monte Carlo simulation, we demonstrate that the proton charge radius can be measured at the 1\% level with several months of data collection using an experimental setup where a photon beam from gamma-ray sources with a flux of $10^7$~photons/s impinges on a TPC active target with a length of 1~meter.
Such a measurement will shed light on the proton charge radius problem and on the lepton universality of the Standard Model.

\bigskip

\acknowledgments 
We are grateful to Wei-Zhi Xiong and Hai-Qing Zhou for useful discussions and Wei-Zhi Xiong for a careful reading of this manuscript.
YHL and UGM are grateful to the hospitality of the Institute of Theoretical Physics, Chinese Academy of Sciences (CAS), where part of the work was done.
This work is supported in part by the CAS under Grants No.~YSBR-101 and No.~XDB34030000;
by the NSFC under Grants No. 12125507 and No. 12047503; and by CAS through the President’s International
Fellowship Initiative (PIFI) under Grant No. 2025PD0022.

\appendix
\section{Differential Cross Section of the Bethe-Heitler Process}\label{app:BHxs}
Here, we provide the explicit expression for the differential cross section of the BH process used in the sensitivity study. After converting the variables from $m_{l\bar l}$ and $\cos\theta^\prime$ to $m_{l\bar l}^2$ and $t$, and integrating out $\cos\theta^*$ and $\phi^*$ in the differential cross section given in Eq.~\eqref{eq:dcs0}, we obtain
\begin{align}
	\frac{d\sigma_{\rm BH}}{d t d m_{l\bar l}^2}=&\,\frac{\alpha^3}{(s-m_p^2)^2}\frac{4\beta}{t^2(m_{l\bar l}^2-t^2)^4}\frac1{1+\tau}\notag\\
	&\times(C_E G_E^2+C_M \tau G_M^2)~,
\end{align}
with $\alpha= e^2/(4\pi)$ the fine-structure constant, $\beta=\sqrt{1-4m_\mu^2/m_{l \bar l}^2}$ the lepton velocity in the $l\bar l$ \textit{c.m.} frame, where the weighting coefficients $C_{E,M}$ multiplying the EMFFs of the proton have the following general structure:
\begin{equation}
	C_{E/M}=C^{(1)}_{E/M}+C^{(2)}_{E/M}\frac1\beta \ln\left(\frac{1+\beta}{1-\beta}\right).
\end{equation}
The coefficients $C^{(1)}_{E/M}$ and $C^{(2)}_{E/M}$ are expressed through Lorentz invariants as
\begin{align}
	C_E^{(1)}&=t(s-m_p^2)(s-m_p^2-\mllb^2+t)(\mllb^4+6\mllb^2 t+t^2+4m_\mu^2\mllb^2)\notag\\
	&\phantom{xx}+(\mllb^2-t)^2[t^2\mllb^2+m_p^2(\mllb^2+t)^2+4m_\mu^2 m_p^2 \mllb^2],\nonumber\\
	C_E^{(2)}&=-t(s-m_p^2)(s-m_p^2-\mllb^2+t)[\mllb^4+t^2+4m_\mu^2\notag\\
	&\phantom{xx}\times(\mllb^2+2t-2m_\mu^2)]+
	(\mllb^2-t)^2[-m_p^2(\mllb^4+t^2)\notag\\
	&\phantom{xx}+2m_\mu^2(-t^2-2 m_p^2 \mllb^2+4m_\mu^2 m_p^2)], \nonumber\\
	C_M^{(1)}&=C_E^1-2m_p^2(1+\tau)(\mllb^2-t)^2(\mllb^4+t^2+4m_\mu^2\mllb^2), \nonumber\\
	C_M^{(2)}&=C_E^2+2m_p^2(1+\tau)(\mllb^2-t)^2\notag\\
	&\phantom{xx}\times[\mllb^4+t^2+4m_\mu^2(\mllb^2-t-2m_\mu^2)].
\end{align}

\section{{Uncertainty of the on-shell approximation to the TCS Process}}\label{app:offshell}

\begin{figure}[t]
	\centering
	\includegraphics*[width=0.48\textwidth,angle=0]{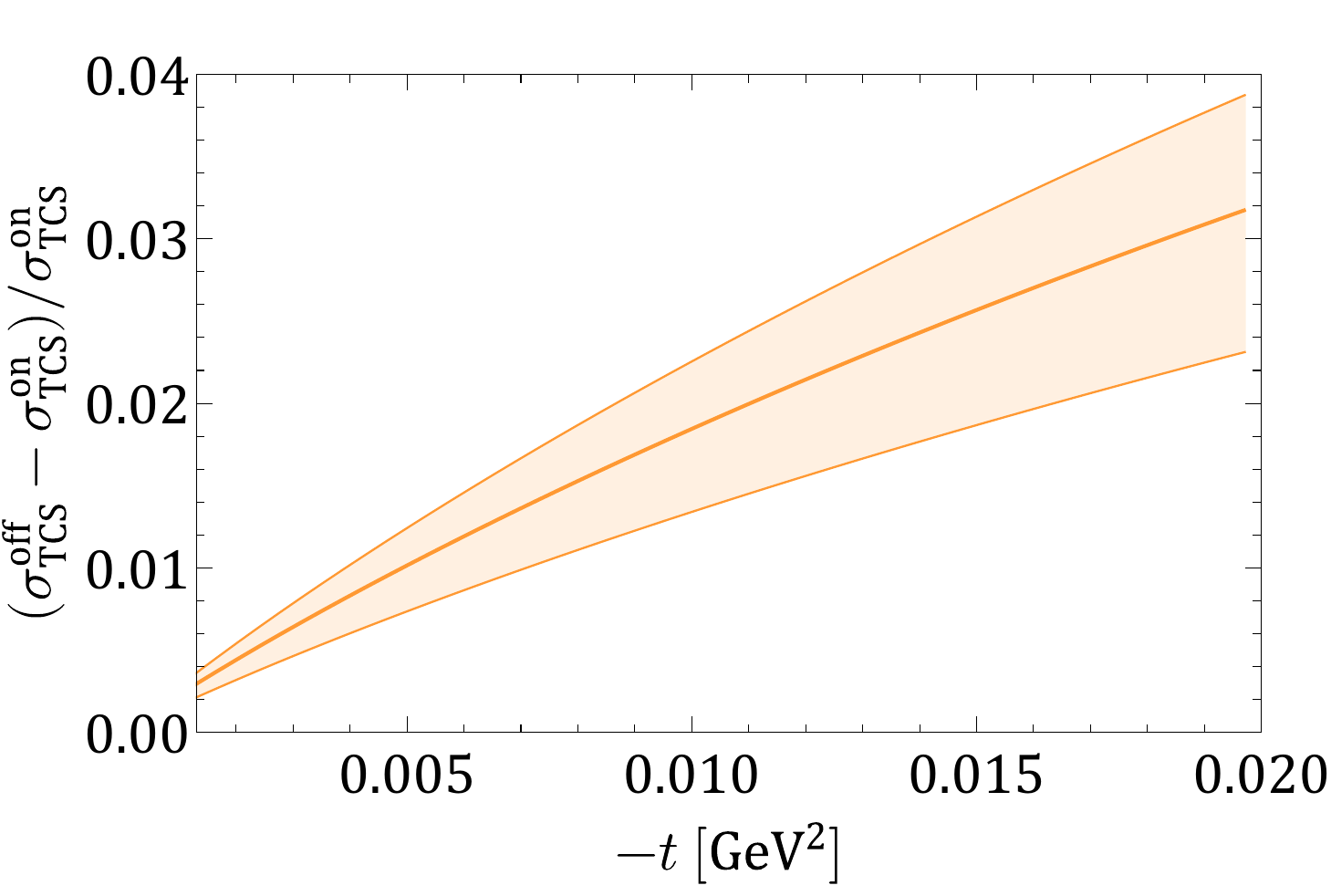}
	\caption{The off-shell contribution to the TCS process that is normalized to the on-shell approximation within the interested kinematic region.
	}
	\label{fig:offshell}
\end{figure}

{The TCS process discussed above involves an intermediate proton that has the four momentum with $p^2\neq m_p^2$. 
The half-on-shell $\gamma^* N^* N$ vertex contains more degrees of freedom than the free nucleon electromagnetic vertex (that is, Eq.~\eqref{eq:onshell}), 
and introduces additional off-shell contributions to the TCS process. 
The off-shell electromagnetic interaction of the nucleon has been intensively investigated in the
literature~\cite{Bincer:1959tz,Naus:1987kv,Bos:1993iz,Haberzettl:1997jg,Koch:2001ii,Haberzettl:2018dpo}. 
We use an extension of the minimal-substitution prescription for coupling the electromagnetic field to hadronic systems proposed in Ref.~\cite{Haberzettl:2018dpo} to estimate the off-shell contributions to the TCS process.

The half-on-shell electromagnetic current operator of the nucleon reads~\cite{Haberzettl:2018dpo}
\begin{align}\label{eq:offshell}
&\Gamma_{\rm off}(p^\prime,p)=\gamma^\mu F_1(q^2)+\frac{i \sigma^{\mu\nu}q_\nu}{2 m_p}F_2(q^2)\notag\\
&\phantom{x}+\frac{q^2}{m_p^3}\frac{p^{\prime 2}-m_p^2}{2 m_p}\bigg[\gamma^\mu D_1(p^{\prime 2},q^2)+\frac{i \sigma^{\mu\nu}q_\nu}{2 m_p}D_2(p^{\prime 2},q^2)\bigg],
\end{align}
where $p$($p^\prime$) is the on(off)-shell momentum of the proton, and $q=p-p^\prime$ denotes the photon transfer momentum. 
The first line is exactly the free $\gamma N N$ operator, while the second line presents the off-shell effects in terms of two finite-difference derivative functions $D_1$ and $D_2$. 
Their definitions are given by
\begin{align}
    D_i(p^{\prime 2},q^2)=2m_p^2 \frac{H_i(p^{\prime 2},q^2)-H_i(m_p^2,q^2)}{p^{\prime 2}-m_p^2},
\end{align}
with $H_i$($i=1,2$) the off-shell electromagnetic form factors of the proton, which are scalar functions of $p^{\prime 2}$ and $q^2$. 
Note that the off-shell terms in Eq.~\eqref{eq:offshell} contribute only when $q^2\neq 0$ by construction.

Due to the absence of any experimental input for these off-shell form factors, we rely on model calculations to estimate the off-shell contributions to the TCS process. Following Ref.~\cite{Haberzettl:2018hcl}, we employ
\begin{equation}
    H_1(p^\prime,q^2)=\frac{\Lambda^4}{\Lambda^4+(p^{\prime 2}-m_p^2)^2} \frac{1}{[1-q^2/(0.71\ {\rm GeV^2})]^2},
\end{equation}
and $H_2=\kappa_p H_1$ for simplicity. Moreover, the phenomenological parameter $\Lambda$ is varied from $0.8$~GeV to $1.2$~GeV to examine the model dependence. We present the numerical results in Fig.~\ref{fig:offshell}, where the off-shell contribution to the TCS process, normalized by the on-shell approximation, is shown. It turns out that the off-shell contribution is positively proportional to the photon momentum transfer squared within the kinematic region considered in this work and is at most around 4\% of the on-shell TCS contribution, i.e., four to five orders of magnitude smaller than the BH contribution. 
Therefore, one can safely drop the off-shell correction to the TCS process.
}

\bibliography{refs}

\end{document}